\documentclass[final,5p,times,twocolumn]{elsarticle} 

\usepackage{graphics}
\usepackage{graphicx}

\usepackage{amssymb}

\journal{Nuclear Instruments and Methods A}

\begin{document}

\begin{frontmatter}



\title{The next generation Cherenkov Telescope Array observatory: CTA}

\author[A]{S.~Vercellone}
\author[B]{for the CTA Consortium} 
\address[A]{INAF/IASF, Via U. La Malfa 153, 90146 Palermo - Italy}
\address[B]{\texttt{http://www.cta-observatory.org/}}

\fntext[fn1]{We gratefully acknowledge support from the agencies and organizations 
listed in this page: \texttt{http://www.cta-observatory.org/?q=node/22}}

\begin{abstract}

The Cherenkov Telescope Array (CTA) is a large collaborative effort
aimed at the design and operation of an observatory dedicated to the
very high-energy gamma-ray astrophysics in the energy range 30 GeV--100 TeV,
which will improve by about one order of magnitude the sensitivity with 
respect to the current major arrays (H.E.S.S., MAGIC, and VERITAS).
In order to achieve such improved performance, for both the northern and 
southern CTA sites, four units of 23\,m diameter Large Size Telescopes (LSTs) 
will be deployed close to the centre of the array with telescopes 
separated by about 100\,m. A larger number (about 25 units)
of 12\,m Medium Size Telescopes (MSTs, separated 
by about 150m), will cover a larger area. The southern site will also include
up to 24 Schwarzschild-Couder dual-mirror 
medium-size Telescopes (SCTs) with the primary mirror diameter of 9.5\,m.

Above a few TeV, the Cherenkov light intensity is such that showers can
be detected even well outside the light pool by telescopes significantly
smaller than the MSTs. To achieve the required sensitivity at high
energies, a huge area on the ground needs to be covered by Small Size Telescopes 
(SSTs) with a field of view of about 10 degrees and an angular resolution 
of about 0.2 degrees, making the dual-mirror configuration very effective. 
The SST sub-array will be composed of 50--70 telescopes with a mirror area of about 
5--10 square meters and about 300\,m spacing, distributed across an area of about 
10 square kilometers.

In this presentation we will focus on the innovative solution for the optical design
of the medium and small size telescopes based on a dual-mirror
configuration. This layout will allow us to reduce the dimension and the
weight of the camera at the focal plane of the telescope, to adopt
Silicon-based photo-multipliers as light detectors thanks to the reduced
plate-scale, and to have an optimal imaging resolution on a wide field of view.

\end{abstract}


\end{frontmatter}

	\section{Introduction}

The very high-energy (VHE) portion of the electromagnetic spectrum (above $\approx 100$\,GeV)
is currently being investigated by means of ground-based imaging array Cherenkov telescopes (IACTs). 
The telescope optical system adopted by the current IACTs (H.E.S.S., MAGIC, and VERITAS) is composed of a 
tessellated mirror (D$\approx 12-17$\,m) which focusses the Cherenkov light on a focal plane
covered by photomultipliers. The stereoscopic approach allowed to substantially improve the
background rejection, the energy/angular resolution and yield the discovery of more than 150 sources,
among Galactic, extragalactic and unidentified one (see J.~A.~Hinton and W.~Hofmann~\cite{HH-09} for a recent
review). 

In order to dramatically boost the current IACTs performance and to widen the VHE science, a new 
Cherenkov telescope array (CTA) has been proposed, as described in  M.~Actis~\cite{A-11} and more
recently in B.~S.~Acharya~\cite{A-13} and in the CTA Consortium contributions to the 
$33^{rd}$ ICRC Symposium~\cite{CTA-ICRC13}.
CTA plans the construction of many tens of telescopes divided in three kinds of configurations.
Two arrays will be deployed (construction starting from 2015), one in the northern and one in the southern hemisphere, 
in order to provide all-sky coverage. The wide energy range covered by the CTA (30\,GeV--100\,TeV) requires
different kinds of telescopes. We plan to have 4 large size-telescopes (LSTs, D$\sim23$\,m) at the center of the array
(to lower the energy threshold down to E$\sim30$\,GeV), 25 medium size-telescopes (MSTs, D$\sim12$\,m) covering 
about 1\,km$^{2}$ (to improve by a factor of ten the sensitivity in the energy range 0.1--10\,TeV).
Moreover, --in the southern site--  we plan to install 50-70 small size telescopes 
(SSTs, primary mirror D$\sim4$\,m, Aeff$\sim5-10$\,m$^{2}$) covering about 10\,km$^{2}$
(to extend Galactic plane source studies in the energy range beyond 100 TeV),
in conjunction with 24 Schwarzschild-Couder dual-mirror telescopes
(SCTs, primary mirror D$\sim9.5$\,m) to further improve the angular resolution in the energy range 0.1--10\,TeV.

The extremely wide energy band, the one-order-of-magnitude improvement in the overall sensitivity, the optimal
angular ($0.1^{\circ}$ at 0.1\,TeV; $0.05^{\circ}$ at 1\,TeV) and energy ($\Delta E/E \le 25$\% at 50\,GeV; 
$\Delta E/E \le 10$\% at 1\,TeV) resolution will allow us to address the scientific topics in a two-fold approach.
From one side, CTA will investigate a much larger number of already known classes of sources, going to much
larger distances in the Universe, performing population studies, accurate variability and spatially-resolved studies.
On the other side, such performance figures will allow new light  to be shed on possible new classes of TeV sources,
such as GRBs, cluster of Galaxies, Galactic binaries, and address fundamental physics studies. Moreover, pushing
the high-energy limit to E$\ge 100$\,TeV will allow a thorough exploration of the cut-off regime of the cosmic accelerators.
CTA will be operated as an Observatory, open to the scientific community by means of peer-reviewed announcement
of opportunity observations.

	\section{The one-mirror telescopes}
\begin{figure}[htb]
\centerline{
\includegraphics[width=0.5\columnwidth,angle=270]{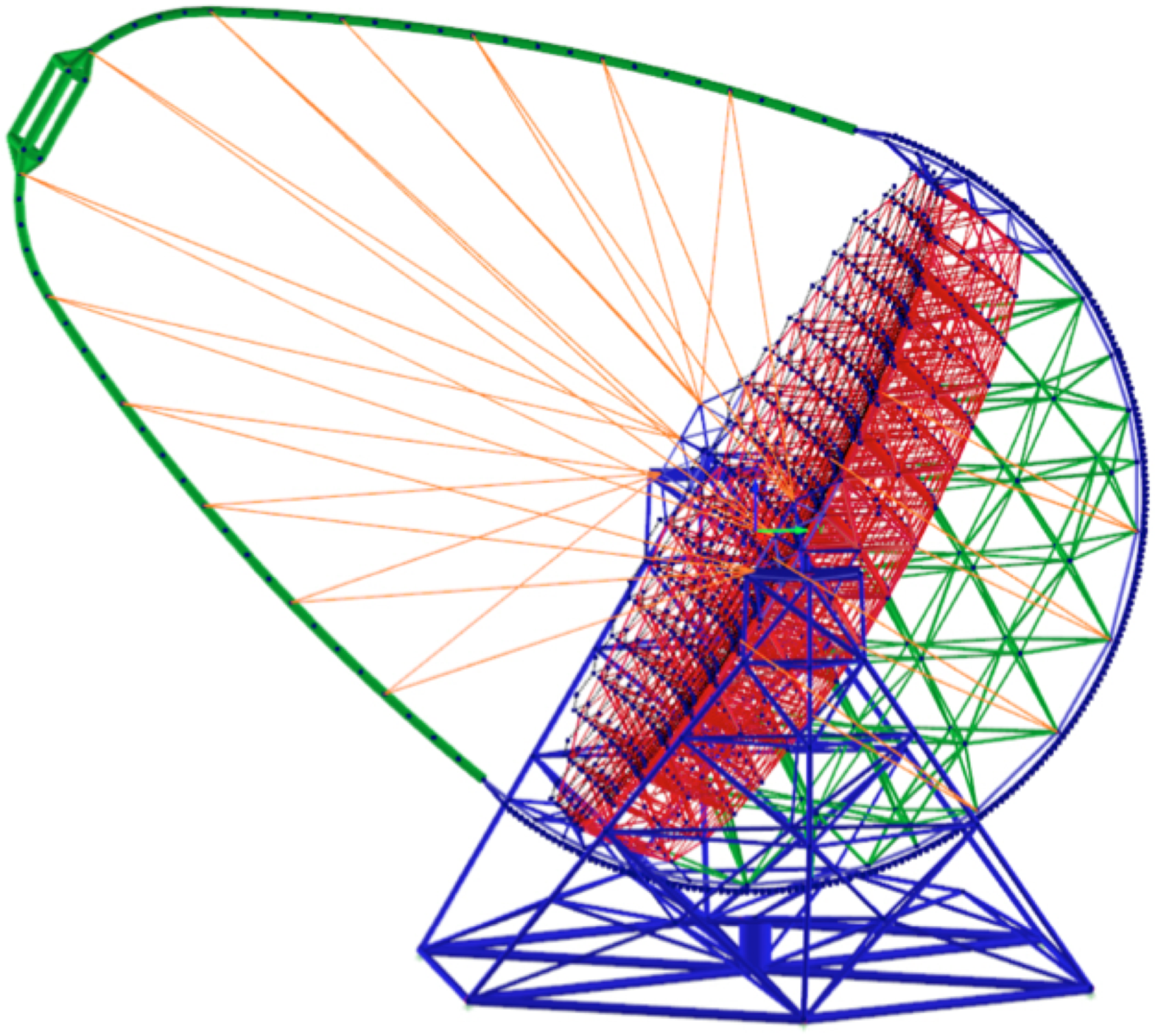}}
\caption{Basic design of the large size telescope structure.}
\label{fig1}
\end{figure}

Currently, one-mirror telescopes are under study for all the CTA telescope sizes.\\
%
%
{\it \underline{The large-size telescope.}} LSTs (see G.~Ambrosi~\cite{A-ICRC13} for a recent review) 
will provide their contribution in the lower energy range (30--a few hundreds\,GeV).
The science topics addressed by the LSTs require a moderately large field of view (FoV) to study 
extended sources, a fast repointing system to rapidly follow-up GRB triggers and a long focal distance to optimise the
optical performance of the telescope. Fig.~\ref{fig1} shows a 3D rendering of the LST. 
The telescope will adopt a 23\,m dish with a parabolic profile, composed of 198 
hexagonal mirrors, a focal length F=28\,m, a field of view FoV=4.5$^{\circ}$, and
a ratio F/D=1.2. The focal plane will be composed of about 2500 photo-multipliers (PMTs), 
with a pixel size of 0.1$^{\circ}$. 
The telescope will be able to rotate to any point in the sky above $30^{\circ}$ in elevation in at most 50\,s.

%
%
{\it \underline{The medium-size telescope.}} MSTs (see B.~Behera~\cite{B-SPIE12} for a recent review)
will be devoted to the study of the sources in the core energy range 
0.1--10\,TeV. By reaching milliCrab sensitivity in this energy range, it will be possible to perform population studies,
to investigate extended sources, and to perform accurate variability studies.
In order to obtain an order of magnitude improvement with respect to the current arrays, both the collecting area
and the image reconstruction have to be improved. A much larger number of telescope units with respect to the current
one (four) will accomplish both tasks. 
\begin{figure}[htb]
\centerline{
\includegraphics[width=0.5\columnwidth,angle=270]{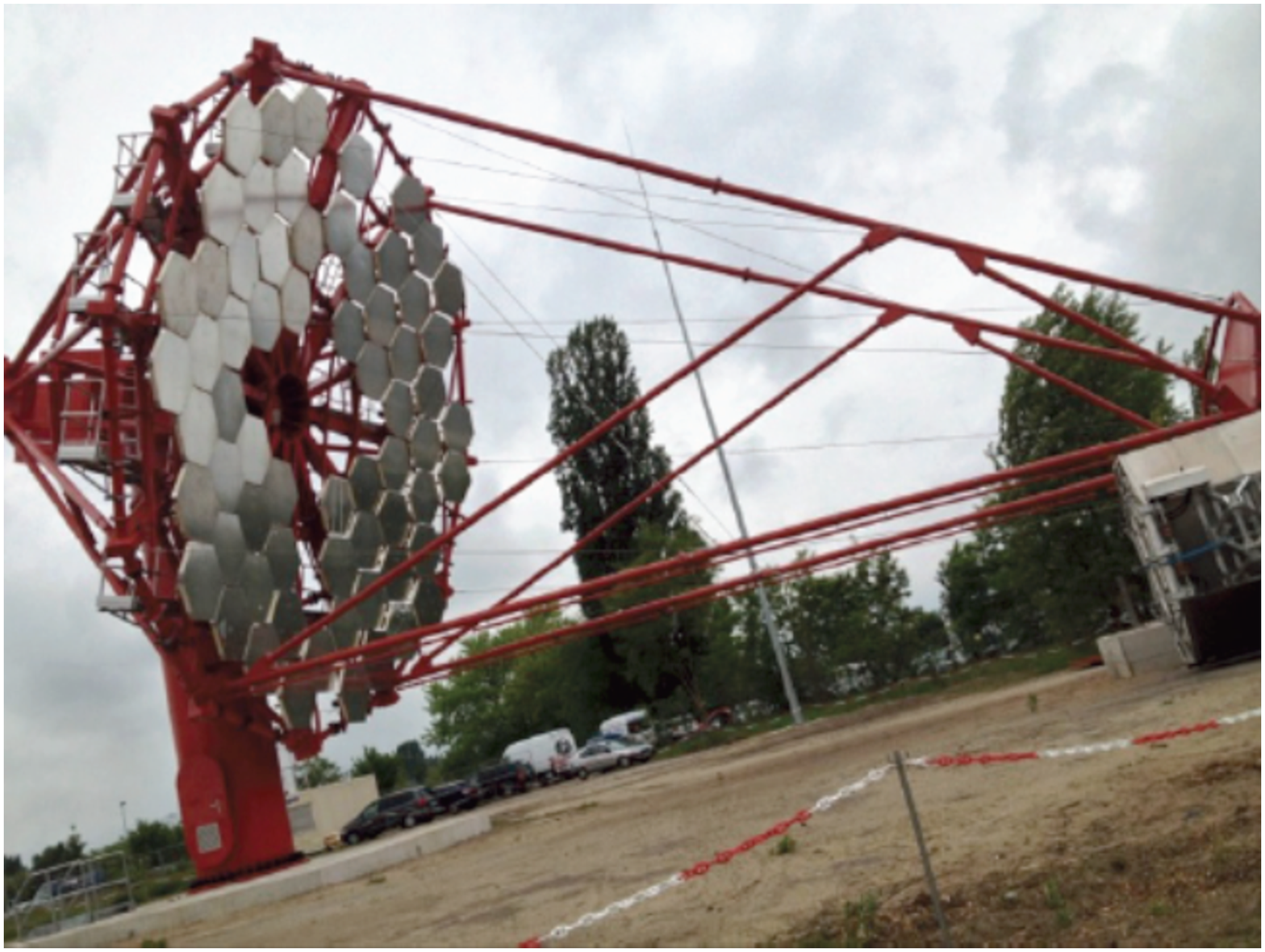}}
\caption{The MST full-scale prototype.}
\label{fig2}
\end{figure}
Fig.~\ref{fig2} shows a 1:1 prototype of the MST installed at the DESY Facility in Berlin. 
The full-scale prototype of the MST structure includes drive and safety system, a dummy-camera, and a mix of 25 real 
and dummy mirrors. The telescope will adopt a 12\,m dish with a Davies-Cotton profile, composed of 86 
hexagonal facets, a focal length F=16\,m, a field of view FoV$\sim7-8^{\circ}$, and
a ratio F/D=1.3. The focal plane will be composed of about 1800 PMTs, 
with a pixel size of 0.18$^{\circ}$. 
The telescope will be able to rotate to any point in the sky above $30^{\circ}$ in elevation in at most 90\,s.

\begin{figure}[htb]
\centerline{
\includegraphics[width=0.5\columnwidth,angle=0]{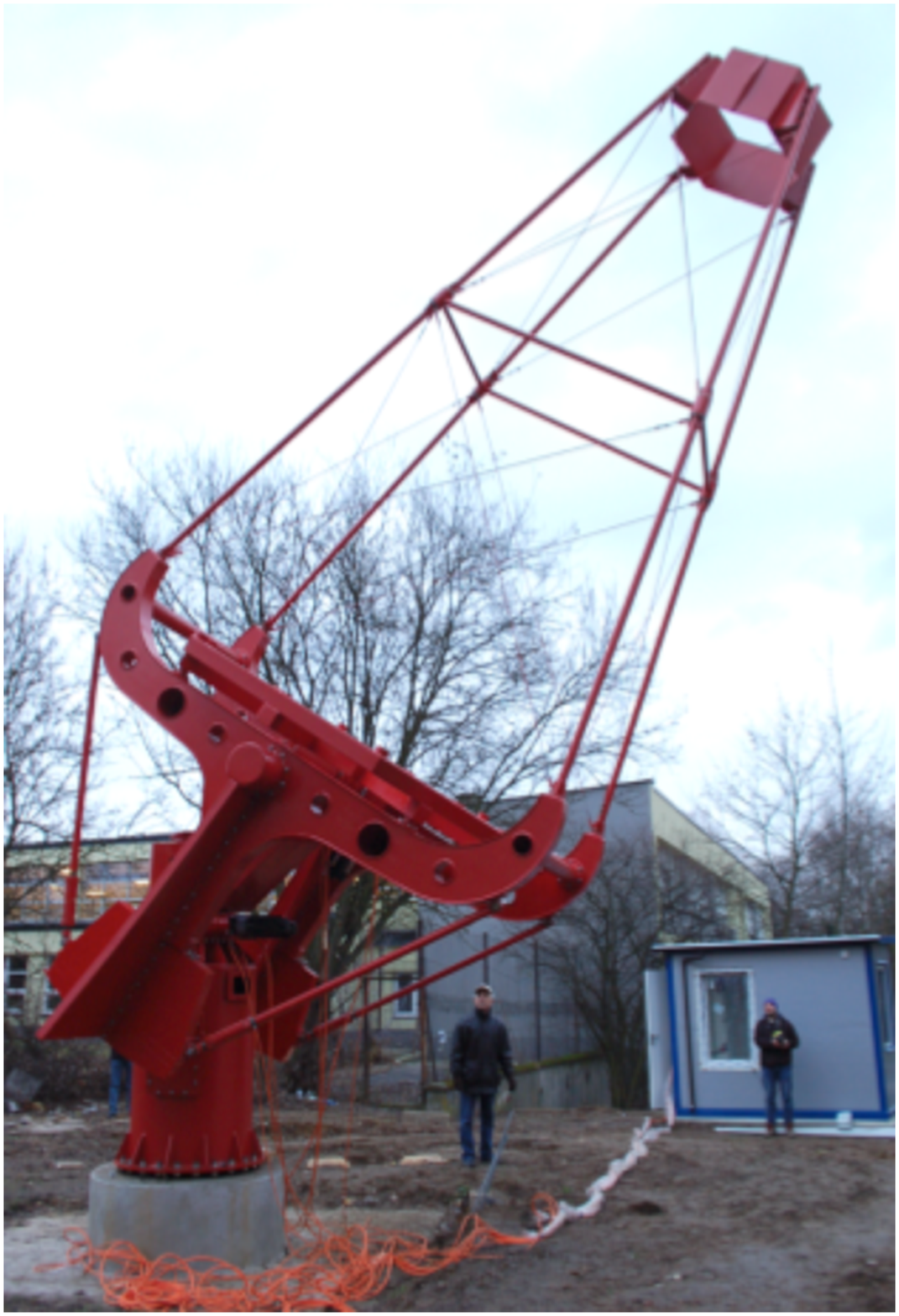}}
\caption{The SST-1M full-scale structural prototype.}
\label{fig2b}
\end{figure}

%
%
{\it \underline{The one-mirror small-size telescope.}} SST-1M (see R.~Moderski~\cite{M-ICRC13} for a recent review)
has the goal to build and install an SST one-mirror prototype in Poland.
Fig.~\ref{fig2b} shows full-scale structural prototype installed at the IFJ site in Krakow.
The telescope will adopt a segmented 3.98\,m Davies-Cotton mirror composed of 18 hexagonal segments. It will have 
a focal length F=5.6\,m, a field of view FoV$\sim9^{\circ}$, and a ratio F/D$_{1}$=1.4. 
The focal plane will be composed of about 1296 hexagonal-shape Geiger-avalanche photo-diodes (G-APDs), 
with a pixel size of 0.25$^{\circ}$. 

	\section{The Schwarzschild-Couder Optical Design}
To further improve CTA performance for the widest spectrum of science topics it is mandatory to
improve precision performance and collection area in the central energy domain (around 1 TeV) and
at the highest energies (above 10\,TeV and up to 100\, TeV).
A possible solution is to increase the number telescopes, the telescope field of view, and to improve the angular resolution.
Nevertheless, small pixel size, large field of view, and controlled cost requirements are mutually
incompatible within the Davies-Cotton telescope design paradigm.
A new dual-mirror, Schwarzchild-Couder (SC) based aplanatic design has been
proposed and developed by V.~V.~Vassiliev, S.~Fegan \& P.~Brousseau~\cite{VFB-07}.
In the SC telescope, the focal plane is located in-between two aspherical mirrors, close to the secondary mirror.
No Cherenkov telescope adopted this optical system up to now.

The dual-mirror optical system will reduce the dimension, the weight, and the cost of the camera
at the focal plane of the telescope, and will obtain a more compact and stiffer mechanical structure, and
 an optimal imaging resolution across a wide field of view.
Moreover, thanks to the reduced plate-scale, silicon-based photo-multipliers (SiPMs) can be adopted as light detectors.
Among other advantages,  it has been demonstrated (A.~Biland~\cite{B-14}) that SiPMs allow to perform observations 
during moonlight, increasing the observatory duty-cycle.

	\subsection{The medium-size Schwarzschild-Couder telescope}
\begin{figure}[htb]
\centerline{
\includegraphics[width=0.57\columnwidth,angle=0]{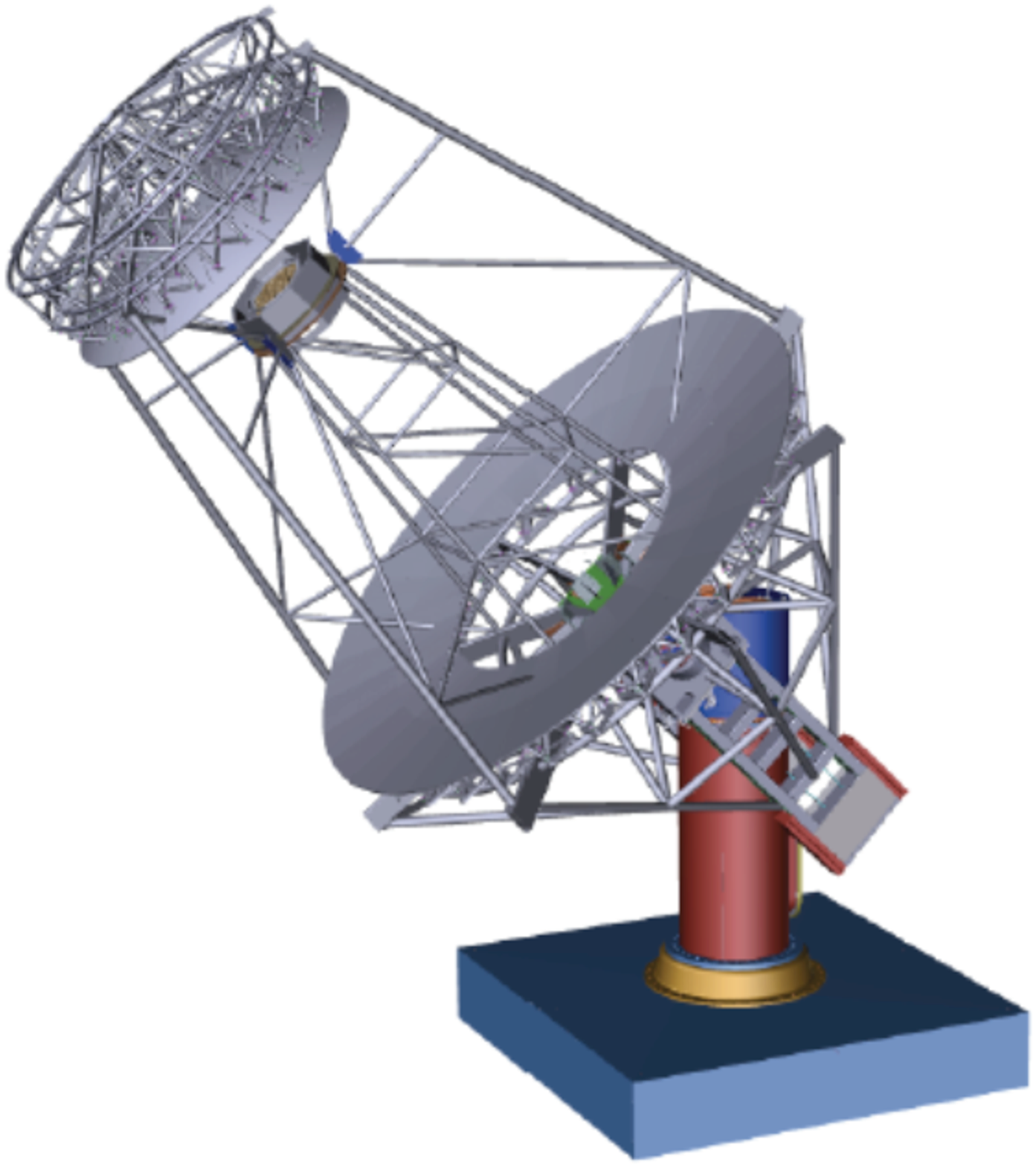}}
\caption{The SCT basic design concept.}
\label{fig3}
\end{figure}

The medium-size SCT (see J.~Rousselle~\cite{R-ICRC13} for a recent review) will constitute the U.S.
contribution to the CTA project.
Fig.~\ref{fig3} shows the basic design of the SCT prototype that is planned to be installed at the Fred
Lawrence Whipple Observatory in Arizona.
The telescope will adopt a 9.66\,m primary mirror (M1) composed of 48 aspherical segments, a 5.42\,m
secondary mirror (M2, 24 aspherical segments), a focal length F=5.59\,m, a field of view FoV$\sim8^{\circ}$, and
a ratio F/D$_{1}$=0.68. 
The focal plane is a novelty as well. It will adopt silicon-based photomultipliers and it will be composed of 
11328 pixels.
The pixel size of 0.067$^{\circ}$ will allow us to approach the physics limit of imaging atmospheric Cherenkov technique. 
The operational energy range combined with the unmatched  angular resolution across the wide field of view 
will allow us to perform accurate spatially-resolved study of extended sources (e.g., supernovae remnants, 
pulsar-wind nebulae).

	\subsection{The small-size Schwarzschild-Couder telescopes}
\begin{figure}[htb]
\centerline{
\includegraphics[width=0.55\columnwidth,angle=0]{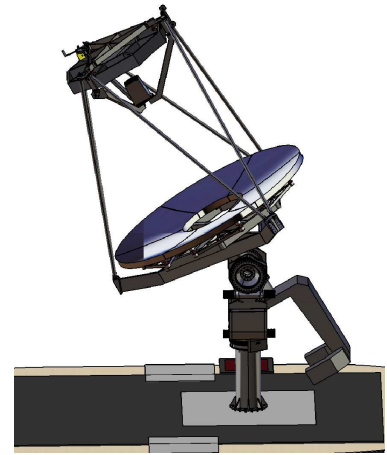}}
\caption{The GATE design concept.}
\label{fig4}
\end{figure}

The CTA-SSTs sub-array (see G.~Pareschi\cite{P-ICRC13} for recent review) is devoted to the exploration 
of the phenomena occurring at the highest energy, above a few TeV. 
At this energy,  the sensitivity is limited by the gamma-ray count-rate, thus it is required to deploy
several (50--70) telescopes of small diameter ($\sim 4$\,m) at a relative distance of about 200--300\,m, and 
covering an area of several square km. The SST sub-array will cover energies up to E$\ge 100$\,TeV,
exploring a poorly-known energy window, both for Galactic and extra-galactic studies. Supernovae remnants,
pulsar-wind nebulae, binary-star systems as well as extreme BL Lac objects and the investigation of the 
near-infrared component of the extra-galactic background  light are some of the possible science topics
to be addressed by the SST sub-array.
%
%
Currently, there are two different designs for the SST structures and cameras.\\
\underline{\it The GATE project.} The Gamma-ray Telescope Elements (GATE, see A.~Zech~\cite{Z-ICRC13} for a recent review) 
project has the goal to build and install an SST prototype at the Paris Observatory in Meudon.
Fig.~\ref{fig4} shows the basic design of the GATE prototype.
The telescope will adopt a segmented 4\,m primary mirror (M1) composed of 6 petals, an assembled-panels 2\,m
secondary mirror (M2, 9 facets), a focal length F=2.28\,m, a field of view FoV$\sim9^{\circ}$, and
a ratio F/D$_{1}$=0.57. 

\begin{figure}[htb]
\centerline{
\includegraphics[width=0.55\columnwidth,angle=270]{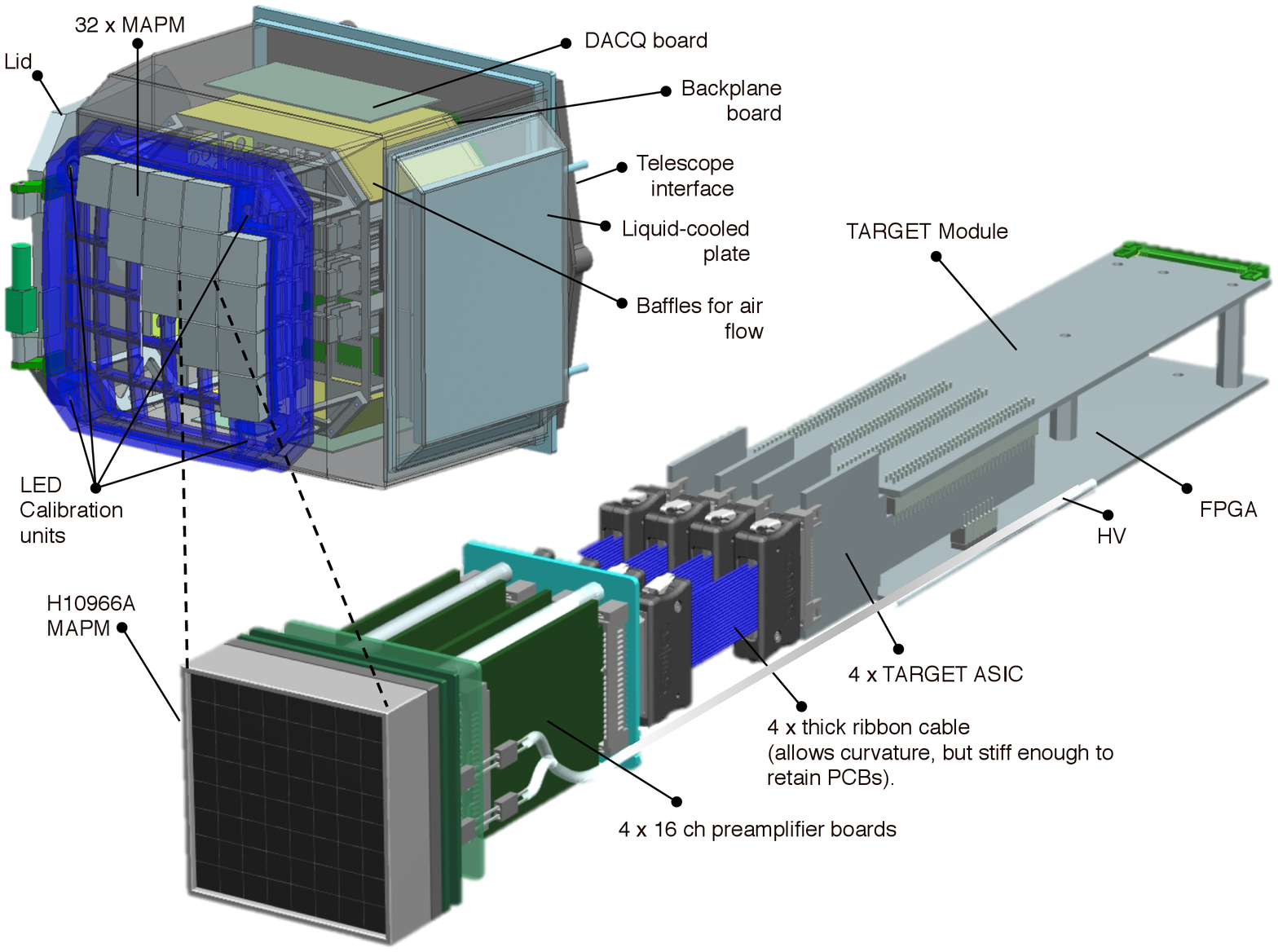}}
\caption{The CHEC design concept (upper left), and the TARGET module (lower right).}
\label{fig5}
\end{figure}

%
%
\underline{\it The CHEC camera.} The Compact High-Energy Camera (CHEC, M.~K.~Daniel~\cite{D-ICRC13}) 
project will be installed on the dual-mirror SSTs, and its basic design and components are shown in Fig.~\ref{fig5}.
A first CHEC prototype (CHEC-M) will adopt  2048 multi-anode photomultipliers with a pixel size of 0.17$^{\circ}$.
The photosensors will be read by a TARGET-5 signal-sampler ASIC. 
A future development (CHEC-S) will consist of a focal plane adopting silicon-based photomultipliers.
The CHEC camera can be installed on both the GATE and the ASTRI telescopes. Moreover, it shares some
components and technological aspects with the SCT camera, such as the front-end board and the backplane. 

%
%
\underline{\it The ASTRI project.} The "Astronomia con Specchi a Tecnologia Replicante Italiana" 
(ASTRI, see the ASTRI contributions to the $33^{rd}$ ICRC Symposium~\cite{ASTRI-ICRC13} for a recent review)
project has the goal of install a fully-functional, end-to-end, dual-mirror SST prototype, and operate it
at the INAF observing station on Mt. Etna (Sicily). 
\begin{figure}[htb]
\centerline{
\includegraphics[width=0.55\columnwidth,angle=0]{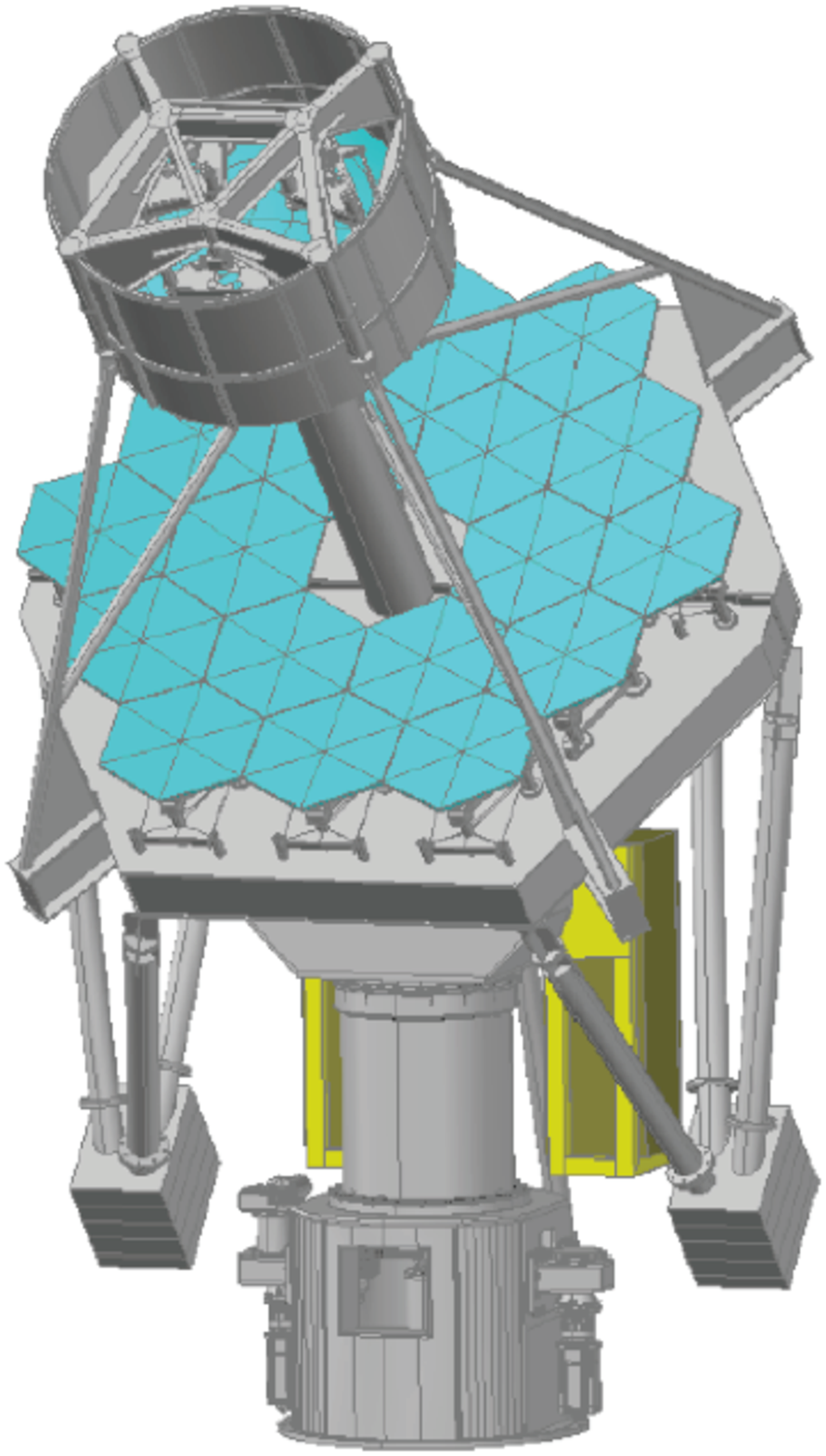}
}
\caption{The ASTRI end-to-end prototype design concept.}
\label{fig6}
\end{figure}
The end-to-end prototype requires, in addition to all the
hardware components, a complete software chain, from the scheduling of the observations down to the data analysis 
and final data archiving. The observing station altitude (1735\,m a.s..l.) and the end-to-end approach will
allow us to perform observations of the Crab, MKN~501 and MKN~421.
Fig.~\ref{fig6} shows the basic design of the ASTRI prototype structural and mirror components,
while Fig.~\ref{fig7} shows the ASTRI camera with its built-in calibration system.
\begin{figure}[ht]
\centerline{
\includegraphics[width=0.65\columnwidth,angle=270]{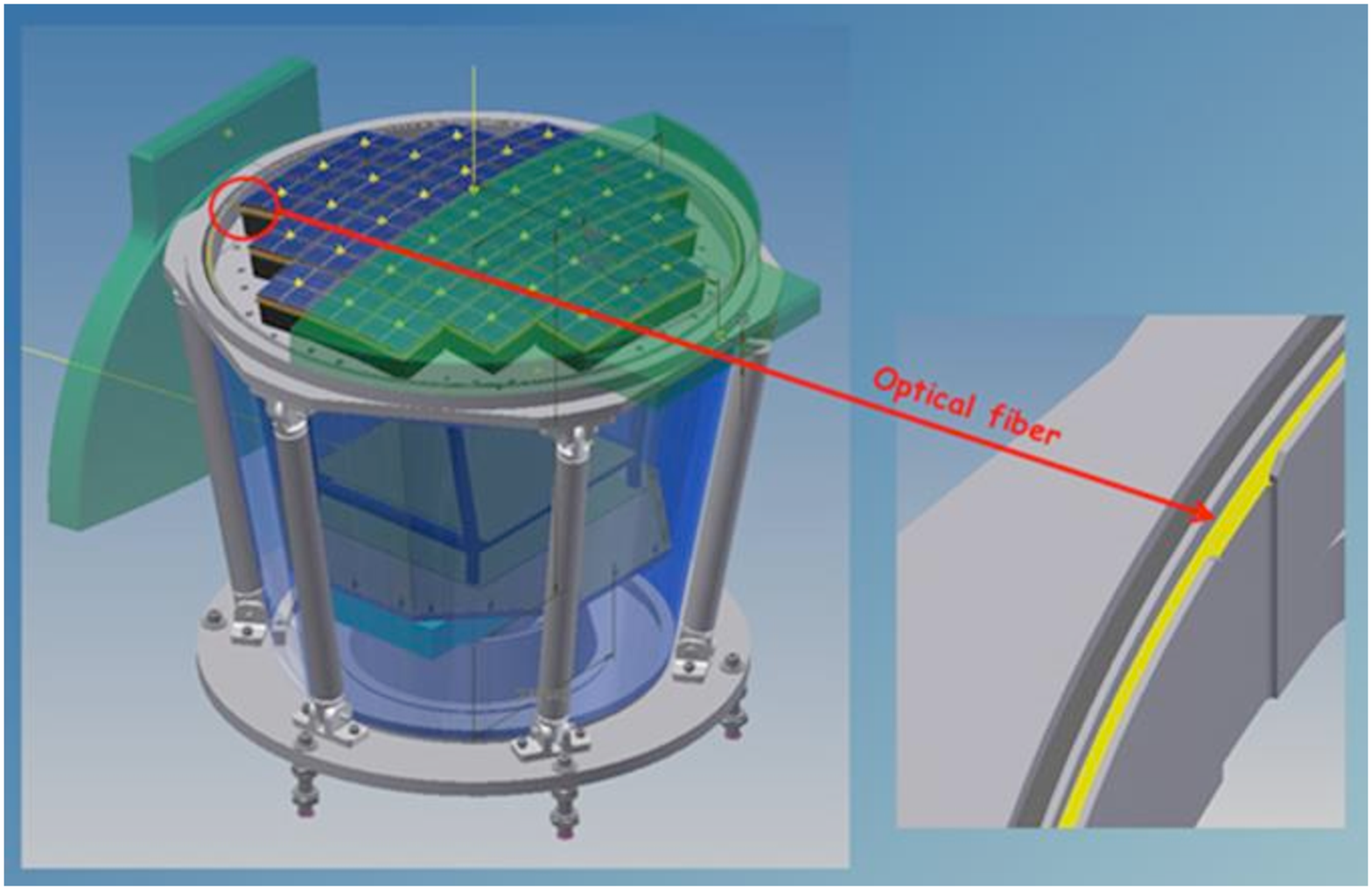}
}
\caption{Basic design of the ASTRI camera. The inset shows the optical fiber system for calibration purposes.}
\label{fig7}
\end{figure}
The ASTRI telescope will adopt a segmented 4.3\,m primary mirror (M1) composed of 18 facets, a monolithic 1.8\,m
secondary mirror (M2, with a radius of curvature of 2.2\,m), a focal length F=2.15\,m, a field of view FoV$\sim9.6^{\circ}$,
for a ratio F/D$_{1}$=0.5. The mirror manufacturing process is the Òglass cold shapingÓ technique, specifically developed 
by INAF for Cherenkov mirrors.
The curved focal plane ($\sim 1$\,m of radius of curvature) will host 1984 6.2\,mm\,$\times$\,6.2\,mm logical pixels
($0.17^{\circ})$. The current photo-sensors are the Hamamatsu S11828-3344M silicon-based photo-multipliers, but
other sensors are under test. The ASTRI camera is extremely compact ($\sim$\,50\,cm\,$\times$\,50\,cm\,$\times$\,50\,cm) and
light ($\sim$\,50\,kg). Contrary to other CTA telescopes, the ASTRI camera will adopt as front-end-electronic the
CITIROC, a customized version of the EASIROC (S.~Callier~\cite{Callier}) ASIC signal shaper manufactured 
by Omega\footnote{\texttt{http://omega.in2p3.fr/}}.

	\section{Conclusions}
CTA represents a major step towards the understanding the very high-energy Universe,
by means of a 10-fold improvement in sensitivity, an analogous to the advance from CGRO/EGRET to {\it Fermi}-LAT.
The new Cherenkov array will both study a much larger sample of known high-energy sources, and at the same time
it has a huge discovery potential for the physics of Galactic and extragalactic sources, and for Fundamental Physics studies,
as discussed in M.~Persic~\cite{Persic}.
The telescope designs are developed combining both a proven technology (LST, MST, SST-1M) and judicious innovations 
(SCT, SST-2M), by introducing the dual-mirror concept and the silicon-based photo-multipliers in the Cherenkov telescope 
manufacturing. Last but not least, CTA will serve the entire astrophysics community, operating at the end of this decade 
as an open Observatory.

\end{document}